\documentclass[
 aps,
 11pt,
 final,
 notitlepage,
 oneside,
 twocolumn,
 nobibnotes,
 nofootinbib,
 superscriptaddress,
 noshowpacs,
 centertags]
{revtex4}

\def\saoname{Special Astrophysical Observatory,  Russian Academy of Sciences,
              Nizhnii Arkhyz, 369167 Russia}

\def\squareforqed{\hbox{\rlap{$\sqcap$}$\sqcup$}}

\def\sq{\ifmmode\squareforqed\else{\unskip\nobreak\hfil
\penalty50\hskip1em\null\nobreak\hfil\squareforqed
\parfillskip=0pt\finalhyphendemerits=0\endgraf}\fi}

\def\degr{\hbox{$^\circ$}}

\def\arcmin{\hbox{$^\prime$}}

\def\arcsec{\hbox{$^{\prime\prime}$}}

\def\utw{\smash{\rlap{\lower5pt\hbox{$\sim$}}}}

\def\udtw{\smash{\rlap{\lower6pt\hbox{$\approx$}}}}

\def\fm{\hbox{$\,.\!\!^{\rm m}$}}

\def\diameter{{\ifmmode\mathchoice
{\ooalign{\hfil\hbox{$\displaystyle/$}\hfil\crcr
{\hbox{$\displaystyle\mathchar"20D$}}}}
{\ooalign{\hfil\hbox{$\textstyle/$}\hfil\crcr
{\hbox{$\textstyle\mathchar"20D$}}}}
{\ooalign{\hfil\hbox{$\scriptstyle/$}\hfil\crcr
{\hbox{$\scriptstyle\mathchar"20D$}}}}
{\ooalign{\hfil\hbox{$\scriptscriptstyle/$}\hfil\crcr
{\hbox{$\scriptscriptstyle\mathchar"20D$}}}}
\else{\ooalign{\hfil/\hfil\crcr\mathhexbox20D}}%
\fi}}

\newcommand{\aaps}{Astron. and Astrophys. Suppl. }

\newcommand{\mnras}{Monthly Notices Royal Astron. Soc. }

\newcommand{\pasp}{Publ. Astron. Soc. Pacific }

\newcommand{\alet}{Astronomy Letters }

\usepackage[utf8]{inputenc}
\usepackage[russian,english]{babel}
\usepackage{graphicx}
\usepackage{amsmath}
\usepackage{amssymb}

\newcommand{\apjl}{Astrophys.~J. Letters }
\usepackage{hyperref}

\begin{document}



%


\title{STOKES-POLARIMETER FOR 1-METER TELESCOPE}

 \received{October 2, 2020} \revised{November 23, 2020}
 \accepted{November 23, 2020}


\author{\firstname{V.~L.}~\surname{Afanasiev}}
 \email{vafan@sao.ru}
\affiliation{\saoname}

\author{\firstname{E.~S.}~\surname{Shablovinskaya}}
  \email{e.shablie@yandex.com}
\affiliation{\saoname}

\author{\firstname{R.~I.}~\surname{Uklein}}
\affiliation{\saoname}

\author{\firstname{E.~A.}~\surname{Malygin}}
\affiliation{\saoname}


\begin{abstract}
We present the ``StoP'' photometer-polarimeter (Stokes-polarimeter) used for observations with  the 1-m telescope of the
Special Astrophysical Observatory of the Russian Academy of
Sciences since the beginning of 2020. We describe the instrument
and its parameters in observations performed in the photometric
and polarimetric modes. We demonstrate the capabilities of the
instrument through the polarimetry of the blazar S5\,0716+714 and
compare the results with those earlier obtained with the 6-m
telescope.
\end{abstract}

\maketitle


\section{INTRODUCTION}
\label{S:int} Reverberation mapping is currently the only method
operating in the optical range that makes it possible to study
the geometry and kinematics of the central regions of active
galactic nuclei (AGNs), such as Seyfert galaxies, quasars, and BL
Lacertae type objects. Spectroscopic observations, in that case,
require the use of 2--3-m telescopes, and long-term monitoring---a
broad cooperation. With 1-m telescopes spectroscopic methods can
be used only for the brightest AGNs, which have already been well
studied. Extensive monitoring  of the broad line regions (BLR)
with small telescopes can be performed only using filters
covering the required portions of the
spectrum. This was first done by
\citet{CherepLyut73}. Automatic monitoring of AGNs with a set of
narrow-band interference filters has become very popular and is
currently being inducted, e.g., with the 43-cm telescope of Wise
Observatory~\citep{Nunez17}.

Photometric observations in medium-band interference
filters within the framework of the BLR reverberation-mapping
program were started on the Zeiss-1000 telescope \citep{Z1000} of
the Special Astrophysical Observatory of the Russian Academy of
Sciences in January 2018 using MaNGaL \citep{mangal} instrument
and then continued with the MMPP (Multi-Mode
Photometer-Polarimeter) \citep{mmpp, Z1000} developed at the
observatory. We used filters from the available set meant for
studying the spectral energy distribution of active
galaxies\footnote{\url{https://www.sao.ru/hq/lsfvo/devices/scorpio-2/filters_eng.html}}.
A description of the technique of reverberation mapping
observations in interference filters and the first results can be
found in \citet{Ukl19}. It became clear in recent years that
important information about the structure of central AGN regions
can be obtained from BLR reverberation mapping in polarized light
\citep{PolRM} and from high-precision photometric and polarimetric
monitoring of BL~Lac objects\footnote{{Hereafter by
``polarization'' we mean linear polarization of radiation.}}
\citep{0716}. It is therefore highly important not only to perform
spectropolarimetric observations of AGNs, which allow the masses
of supermassive black homes (SMBH) at the centers of galaxies to
be independently determined \citep{AP15}, but also to monitor polarized flux in
broad lines in objects with equatorial scattering on the
dusty tori.

In 2019 we initiated programs of polarimetric observations with
the Zeiss-1000 telescope using  MMPP, which at the time was the
only instrument capable of operating in polarimetric mode. Because
of the use of dichroic polaroid the accuracy of polarimetry proved
to be one order of magnitude lower than required for the task (see
Section~\ref{S:modes}). This was primarily due to non-simultaneous
measurement of the polarized flux of the object at different
polaroid turn angles. Measuring linear polarization with an
accuracy of  0.1--0.2\% requires instruments that can measure
three Stokes parameters simultaneously. We developed such an
instrument as a pilot project in 2012, but had to suspend the work
for a number of reasons. The instrument was initially meant for
monitoring of polarization variations in starlike objects---BL~Lac
type sources and cataclysmic variables in broad photometric bands
using the then novel Andor Neo sCMOS detector \citep{scmos}. Such
a detector must allow fast acquisition of images, which is
necessary for suppressing atmospheric flux scintillation in
polarized light. In 2019 we continued our work on that instrument
and thoroughly upgraded it (we replaced its control system and
used a more efficient detector).

In this paper we describe the optomechanical scheme of the
new ``StoP'' (Stokes-polarimeter) instrument for the Zeiss-1000
telescope and its main specifications. We consider the
specificities of observations in the photometric and polarimetric
modes and report the results of polarimetric observations made
with the 1-m telescope and of the calibration of the  instrument.

\section{INSTRUMENT DESCRIPTION}
\label{S:desc}

\begin{figure*}
\includegraphics[width=16cm]{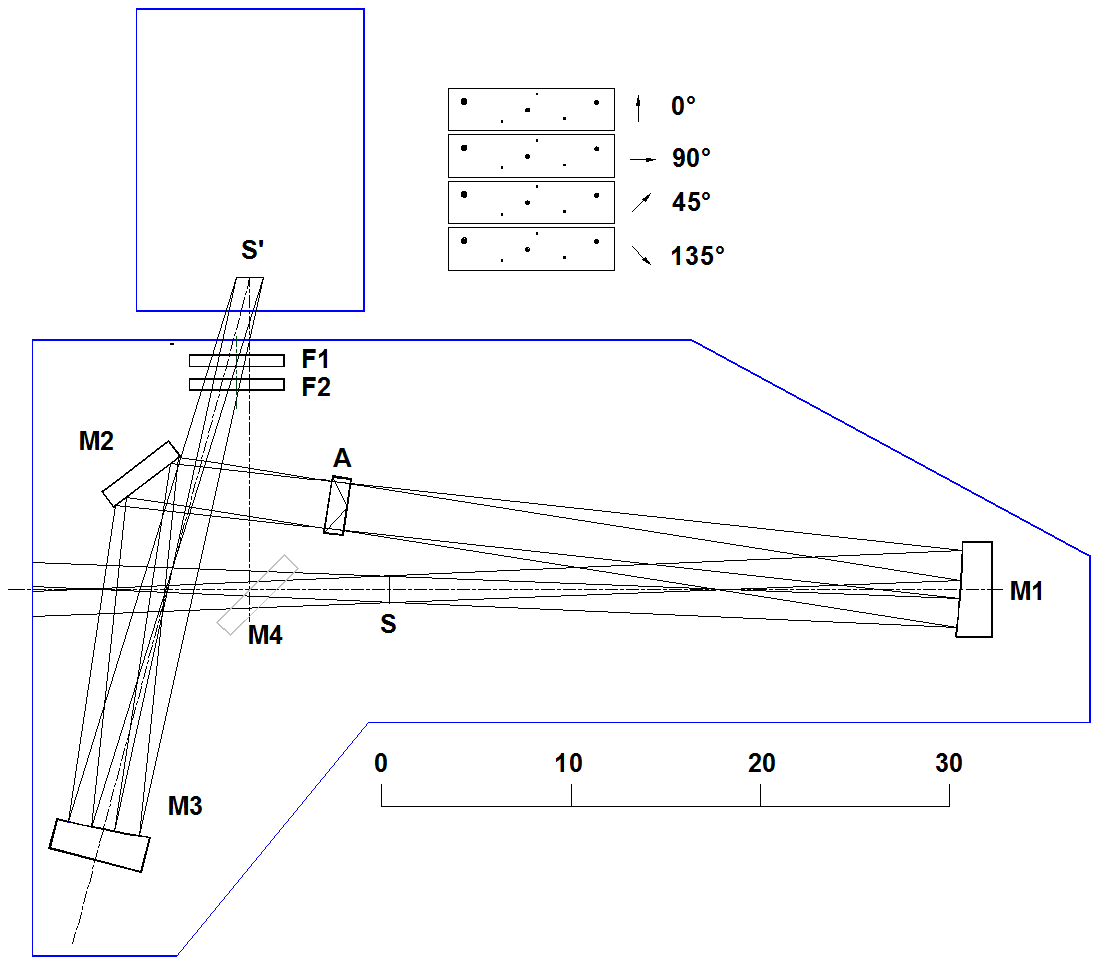}
\caption{Optical layout of the Stokes-polarimeter: M1, M3 are
parabolic off-axis mirrors; M2, M4 are flat diagonal mirrors; A is
the polarization analyzer; F1 and F2 are the turrets with replacement
filters. The scale at  the bottom part of the figure is in
centimeter. } \label{optic}
\end{figure*}

The instrument is meant for operating in two modes: polarimetry
and photometry. The optical layout in the polarimetry mode is made
in accordance with the Czerny-Turner scheme (Fig.~\ref{optic}),
where  focal entrance plane  $S$ is converted into detector plane
$S'$ by using two parabolic off-axis mirrors M1 and M3 with a
focal distance of 300~mm. Polarization analyzer A is mounted at
the entrance pupil of the instrument (the image of the telescope
mirror). The size of the pupil for the entrance focal ratio of
F/12.5 is equal to 24~mm. To reduce the size of the instrument,
the beam is broken by flat mirror M2. We use a double Wollaston prism as the polarization analyzer
\citep{Oliva}.
Such an analyzer consists of two  $30\times 15 \times 5$~mm
Wollaston prisms, which split the pupil into two equal parts,
where each prism separates the ordinary and extraordinary beams by
0.75\degr. To avoid the superposition of  the images produced by
each prism, achromatic wedges are installed behind them that
separate the beams of each prism by 1.5\degr. The first prism is
made so that the ordinary and extraordinary beams correspond to
the electric-vector oscillation directions of 0\degr\ and 90\degr,
respectively, and the corresponding directions for the second
prism are  45\degr\ and 135\degr, respectively. The entrance field
is bounded by a 3.5$\times$24~mm rectangular diaphragm placed
in plane $S$. As a result, as shown in Fig.~\ref{optic}, we obtain
in the detector plane four diaphragm images shifted by 4~mm in
height and corresponding to the electric-vector oscillation
directions of 0\degr, 90\degr, 45\degr\, and 135\degr. The angular
size of the mask-bounded field of view for the 1-m telescope
corresponds to $0.9\times6.1$\arcmin.

Operation in the photometruc mode is provided by placing diagonal
mirror M4 into the beam. Two filter turrets, F1 and F2, each with
9 positions, are placed in front of the output focal plane. The
diameter of the frames of the filters mounted on turrets is equal
to 50~mm, and the allowed thickness cannot be greater than 7~mm.
We use broad-band filters  that we made from {domestic colored glass} and intermediate-band interference filters.

The optical layout of the instrument is meant for the use with
13$\times$18~mm Andor Neo sCMOS detector with a pixel size of
6$\times$6~$\mu$m. Note that the long side of the detector
must be aligned along the direction of the separation of
polarization analyzer beams. However, for a number of
reasons---mostly because of photometric properties in long
exposures and low quantum efficiency, we stopped using this
detector and switched to a low-frame-rate CCD better suited for our task.
The detector installed in our instrument is now
a \mbox{2048$\times$2048}  Andor iKon-L\,936 CCD with a pixel size
of \mbox{13.5$\times$13.5~$\mu$m}, which corresponds to a
scale of  0.21$\arcsec$/pix when working with the 1-m
telescope. However, appreciable geometric vignetting is observed
along the long side of the entrance aperture because of the
specificities of the optical layout. The size of non-vignetted
field of view in that direction is 19.5~mm. Hence the size of
non-vignetted field of view in the polarimetry and photometry
modes is 0.9$\times$5\arcmin and 5$\times$7\arcmin,
respectively. The CCD used in our instrument has a quantum
efficiency greater than 90\% in the 400--850~$\mu$m wavelength
interval and low readout noise. The operating temperature of the
CCD is $-85\degr$C, which ensures that dark noise is lower than
readout noise in 5--10~min exposures. The total weight of the
instrument together with the turning table is equal to 54~kg.

The transmission of the optical system of the instrument is
determined by the reflectivity of mirrors coated with protected
silver,
which amounts to
more than 98\% in the optical range, and by Fresnel losses at the surfaces of Wollaston prisms
which amount to about 7\%. Thus the transmission of the system can be estimated at about
86\% when operating in the polarimetry mode (without taking into
account the quantum efficiency of the CCD, filter transmission curves,
telescope mirror reflectivity).

The control of all mechanical units (switching of filters,
inserting/withdrawing diagonal mirrors, turning the instrument
along the position angle) is provided by the control board with a
microprocessor that is developed to this end. The instrument
incorporates a microcomputer, which sends control commands to the
instrument and to the CCD controller. During observations the
Stokes polarimeter is controlled remotely via  Ethernet using a
program interface written in IDL.

\section{MODES OF OBSERVATIONS}
\label{S:modes}

\subsection{Photometry}

As we pointed out in the previous section, the ``StoP'' instrument
can accommodate up to 16 filters, which can be used for
photometric observations. We use 250\AA-wide medium-band
filters to acquire images and estimate fluxes in broad lines
(mostly Doppler-shifted H$\alpha$ and H$\beta$) and in the
adjacent continuum. Medium-band filters are used to
implement the program of AGN reverberation mapping up to
redshifts  $z \sim 0.8$, and reverberation mapping of Seyfert
galaxies in polarized light.

The set of broad-band glass filters allows implementing the
Johnson-Cousins $UBVR_{\rm c}I_{\rm c}$ photometric
system~\citep{ubvr}. To determine the coefficients for
transforming instrumental magnitudes into the standard photometric
system, we performed on Jaunary 15, 2020 the photometry of stars
in the open cluster  NGC\,2420. {Instrumental magnitudes  $b, v,
r, i$ are computed as \mbox{$-2.5 \log(N)$}, where $N$ is the
number of counts (ADU) from the object in the corresponding band
acquired in the  $1.44 e^-$/ADU gain mode.} To transform
magnitudes to above-atmosphere values, we determined the
extinction coefficients on the same night from observations of the
spectrophotometric standard star Feige\,34. We found the
extinction coefficients to be:
\begin{gather*}
\alpha_B = 0.29^m \pm 0.01^m,\\
\alpha_V = 0.19^m \pm 0.01^m, \\
\alpha_{R_c} = 0.16^m \pm 0.01^m, \\
\alpha_{I_c} = 0.09^m \pm 0.01^m.
\end{gather*}

We selected 16 cluster stars and compared their measured
magnitudes ($bvri$) with published magnitudes ($BVRI$). The
transformation equations have the form:
\begin{gather*}
B = b + 0.08^{\pm 0.03} (B-V) + 22.58^{\pm0.02},\\
V = v - 0.07^{\pm 0.03} (B-V) + 22.85^{\pm0.02}, \\
R = r + 0.11^{\pm 0.07} (V-R) + 22.88^{\pm0.03}, \\
I = i - 0.04^{\pm 0.06} (V-I) + 22.40^{\pm0.05}.
\end{gather*}

\subsection{Polarimetry}

\begin{figure*}
    \includegraphics[width=0.8\linewidth,angle=0]{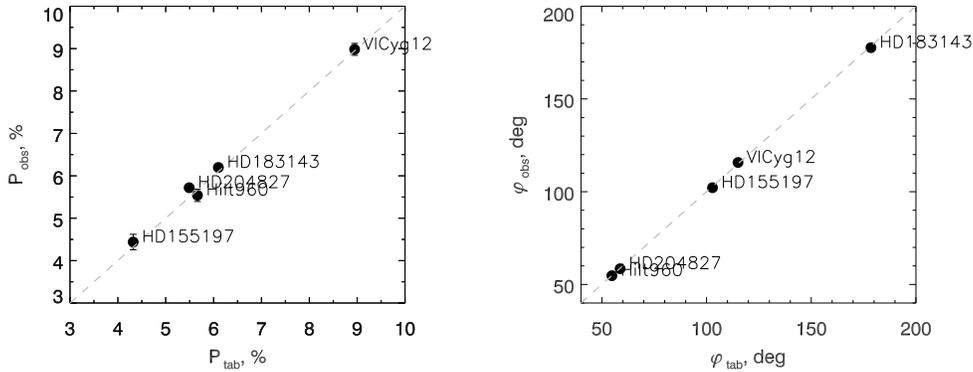}
\caption{Deviation of measured polarization degree values (the
left panel) and polarization plane angles (the right panel) from
published values for the ``StoP'' instrument.} \label{ts}
\end{figure*}

\begin{figure*}
    \includegraphics[width=0.8\linewidth,angle=0]{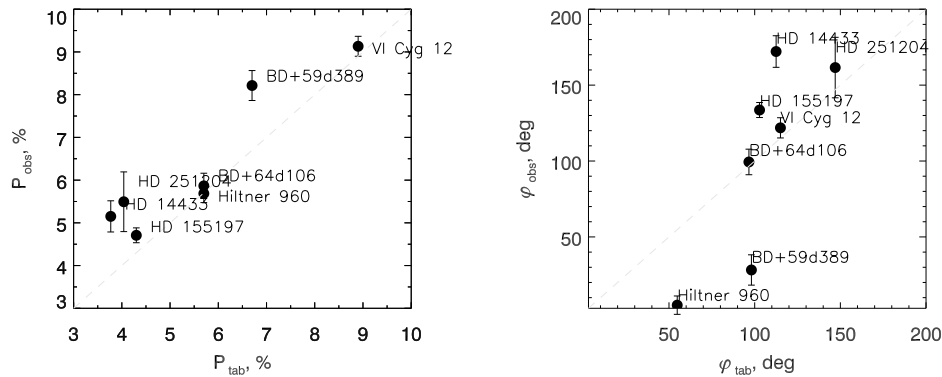}
\caption{Deviation of measured polarization degree values (the
left panel) and polarization angles (the  right panel) from
published values for the  MMPP instrument.} \label{tm}
\end{figure*}

The double Wollaston prism in the  ``StoP'' instrument is used as
the polarization analyzer. Although the prism was proposed for
astronomical observations back in the 1990s
\citep{Oliva,geyer}, it is now rarely used, primarily because the
crystal is difficult to make. The advantage of Wollaston prism is
that it allows one to simultaneously record four polarization
directions---$0^\circ$, $90^\circ$, $45^\circ$, and $135^\circ$,
and hence the three Stokes parameters $I, Q, U$ that describe the
intensity and linear polarization are measured simultaneously.
{The polarization model of the instrument in the form of the
M{\"u}ller matrix for a similar crystal is presented in formula~(3) in
\citet{AfAm}.} The normalized Stokes parameters $Q$ and $U$ can be
computed, up to rotation transformation, by  the following
formulae:
\begin{equation}
   \begin{gathered}
    \frac{Q}{I} = \frac{I_0 - I_{90}D_Q}{I_0 + I_{90}D_Q}, \\
    \frac{U}{I}  = \frac{I_{45} - I_{135}D_U}{I_{45} + I_{135}D_U},
    \end{gathered}
    \label{for}
\end{equation}
where $D_Q$ and $D_U$ are the transmission coefficients of the
polarization channels. These coefficients take into account both
instrumental polarization and atmospheric depolarization
\citep[see][for details]{AfAm}. The polarization degree $P$ and
the angle of the polarization plane $\varphi$ then are equal to:
\begin{gather*}
{P} = \sqrt{Q^2 + U^2},\\
{\varphi'} = \frac{1}{2} \arctan Q/U,
\end{gather*}
where ${\varphi'}$ is the instrumental angle of polarization
plane. To transform it to the true value we use the following
formula:
$$ {\varphi} = {\varphi'} + {\rm PA} + 89.1^\circ,$$
where ${\rm PA}$ is the position angle of the instrument.

Simultaneous registration of fluxes in four polarization
directions allows one to control rapidly varying parameters of
atmospheric depolarization and thereby improve the accuracy of
polarimetry compared to the use of other analyzers. The only big
disadvantage of double Wollaston prism is the need to use a
rectangular mask, which decreases the size of the field of view
several times. That is why prism is used in spectropolarimetric
observations and for the polarimetry of starlike objects like it
is done in the case of SCORPIO-2 instrument at the 6-m telescope of the Special
Astrophysical Observatory of the Russian Academy of Sciences.

Yet another specific feature of the operation in polarimetry mode
is the fact that Wollaston prism has its own dispersion. Chromatic
aberrations introduced by the prism distort images in the
cross-mask direction and have an angular scale of
$\leq$1.5\arcsec, which is comparable with typical seeing at the Special
Astrophysical Observatory. In the case of observations of closely
located objects the instrument can be aligned using the turning table
so as to prevent the superposition of object images broadened by
chromatic aberrations.

The optical system of the instrument uses inclined mirrors,
resulting in instrumental polarization. To measure the degree of
instrumental polarization over several observational spring sets
standard stars with zero polarization were imaged. The zero level of instrumental
polarization was found to be~0.74\% with across-field variations
below~0.1\%. The direction of instrumental polarization coincides
with that of the mask to within~\mbox{$\pm3^\circ$}. Stray
light provides no significant contribution to instrumental
polarization.

Fig.~\ref{ts} compares the results of our observations of
polarized standards to published data. We found the error of
polarization degree and polarization angle to be $({P}_{\rm tab} -
{P}_{\rm obs}) \sim  0.15\%$ and $({\varphi}_{\rm tab} -
{\varphi}_{\rm obs}) \sim  0.7^\circ$, respectively.

For comparison, we show in Fig.~\ref{tm} similar dependences of
the observed polarization degree and polarization angle on
published values for  MMPP instrument \citep{mmpp}, which has been
used on Zeiss-1000 telescope since 2018 and which was operated in
test mode at the time when this paper was written. The
polarization  analyzer employed is a dichroic polaroid mounted in
three positions--- $-60^\circ$, $0^\circ$, and $60^\circ$---and
producing a field of view with a size of about
6$\times$6\arcmin. Because of non-simultaneous imaging of the
object at different polarization directions and slow rotation of
the polaroid (up to 20 seconds between the positions) measurements
performed using this technique become especially sensitive to
weather and their accuracy degrades appreciably. As is evident
from the plots, the errors of the inferred polarization degree and
polarization angle are equal to $({P}_{\rm tab} - {P}_{\rm obs})
\sim  1.1\%$ and $({\varphi}_{\rm tab} - {\varphi}_{\rm obs}) \sim
45^\circ$, respectively. Such accuracy is insufficient for
studying extragalactic objects.

\section{OBSERVATIONS}
\label{S:0716}

The main task of the ``StoP'' instrument is to perform
high-precision polarimetric observations of starlike objects, and
in this sense are unparalleled among attachable equipment for
Zeiss-1000 telescope. The technique of the measurement of the
polarization of point-like sources with an accuracy of  0.1\% was
successfully used on the 6-m telescope of the Special
Astrophysical Observatory of the Russian Academy of Sciences
equipped with  SCORPIO-2 instrument~\citep{0716}, where
observations involved the use of an analyzer similar to that
incorporated in ``StoP''---double Wollaston prism WOLL-2. To test
the technique of observations we decided to repeat observations of
the blazar S5 0716+714.

\begin{figure}
    \centering
    \includegraphics[width=0.9\linewidth]{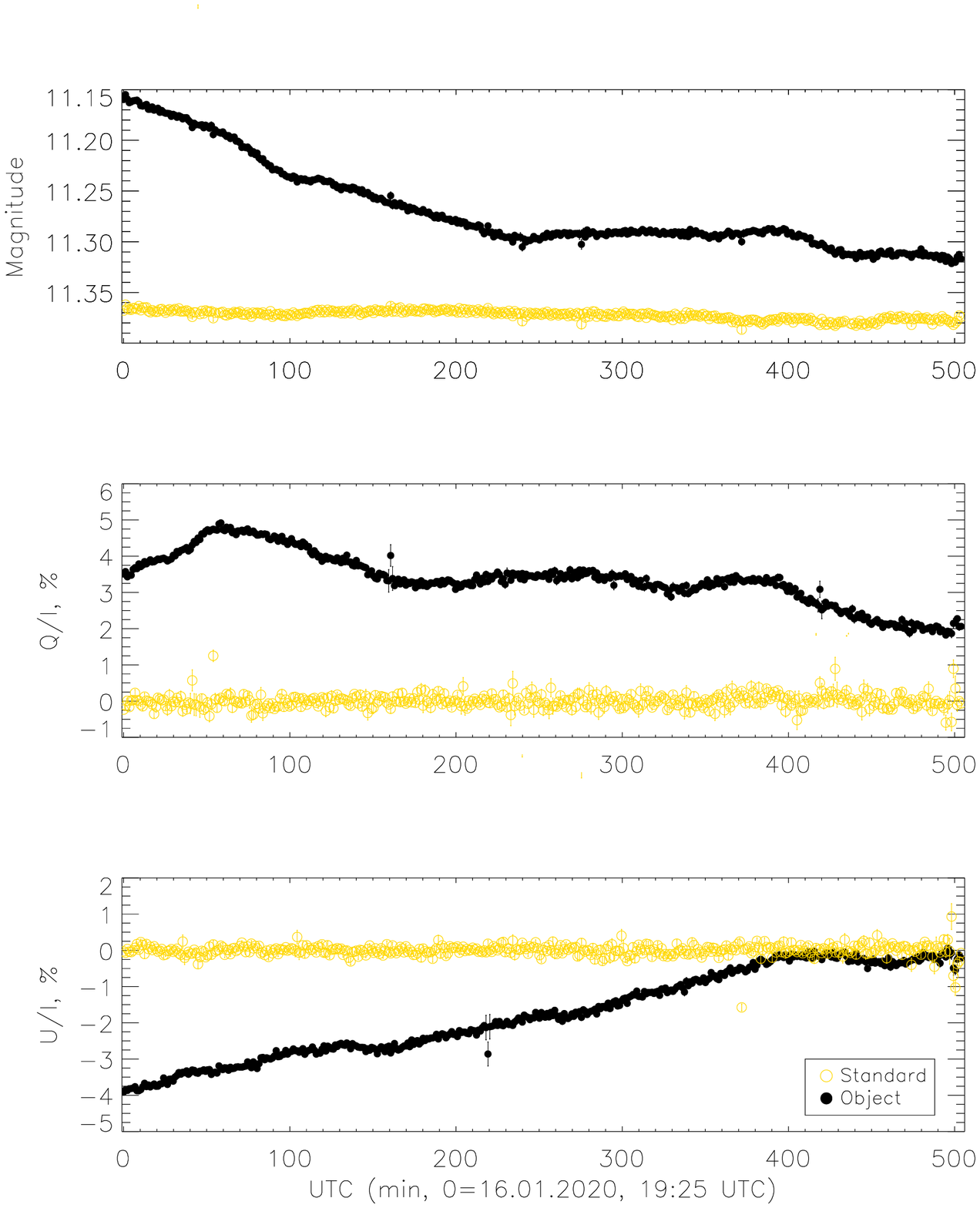}
    \caption{Photometric variations and variations of normalized Stokes parameters $Q$ and $U$ during the night.}
    \label{F:IQU}
\end{figure}

On January 16, 2020 the ``StoP'' instrument attached to Zeiss-1000
telescope was used to perform 8.5-hour monitoring of the blazar
S5\,0716+714 in polarized light. 4400 frames of 60-second
exposure each were acquired separated by readout time no greater than
10 seconds. Observations were made in white light. The mask was
aligned so that the field of view would cover two standard stars
\citep{Amir06} located near the object. The standard stars have
constant flux and zero polarization and therefore were used for
differential polarimetry of the object. We computed the normalized
Stokes parameters by formulae~\eqref{for} relative to reference
stars in the field.

\begin{figure}
    \centering
    \includegraphics[width=0.95\linewidth]{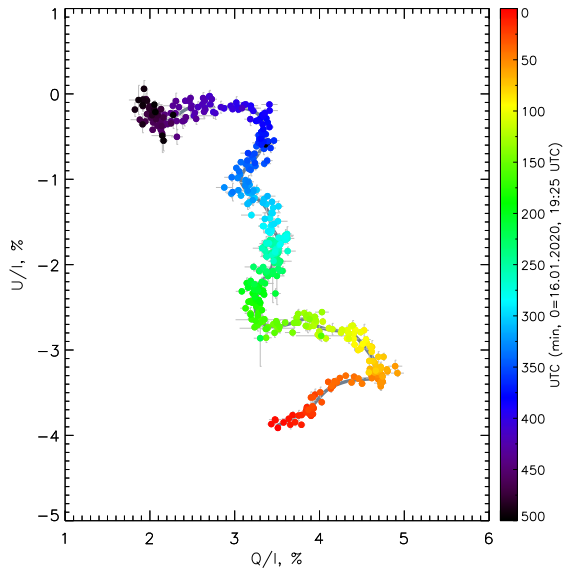}
    \caption{Overnight variations of the normalized Stokes parameters $Q$ and $U$ projected onto the  $QU$-plane.}
    \label{F:QU}
\end{figure}

\begin{figure}
    \centering
    \includegraphics[width=0.95\linewidth]{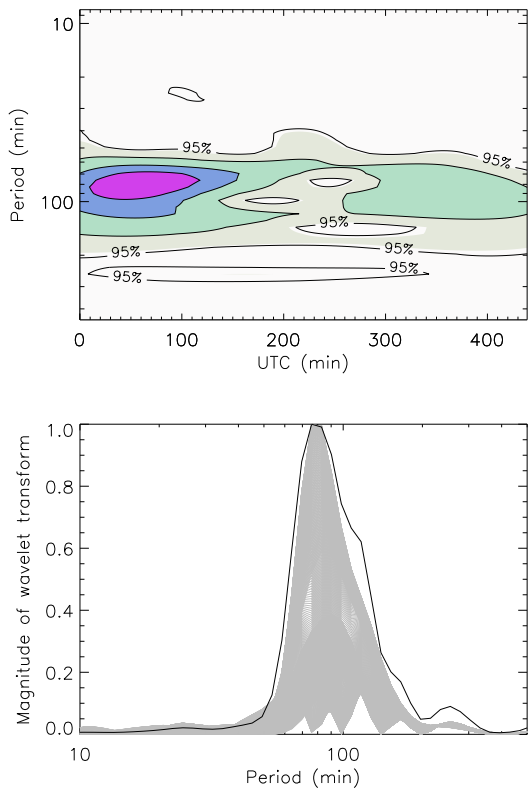}
    \caption{Result of a wavelet analysis of the light curve of the blazar S5 0716+714.
    The maximum of the profile of the wavelet  transform corresponds to the light-variation period of  $\sim$76 minutes.}
    \label{F:wave}
\end{figure}

Fig.~\ref{F:IQU} shows the results of observations. Photometric
errors do not exceed $0\fm005$ and the average error of
polarization measurements is  0.05\%. Observations in
Fig.~\ref{F:QU} are shown in the form of the $QU$-diagram where
the time elapsed since the start of the observation of the object
is coded by color. As is evident from the figure, polarization
vector varies smoothly during the night with its direction
reversing with a period of about 75~minutes. The same period shows
up in the light curve corrected for the long-period trend
approximated using local linear-regression algorithm (LOESS).
Fig.~\ref{F:wave}  shows the result of a wavelet analysis of rapid
flux variations of the object. Wavelet analysis shows the period
of light variations of the object to be \mbox{76$\pm$10}~minutes.
The agreement between the period of light variations and that of
the variations of the polarization vector are totally consistent
with the result that we earlier obtained based on observations
made with the 6-m telescope of the Special Astrophysical
Observatory of the Russian Academy of Sciences \citep{0716}. Thus
the observations mentioned above show that the task of the
polarimetry of relatively bright  (about $13$--$14^{\rm m}$)
starlike objects can be successfully addressed with the  ``StoP''
instrument and the accuracy of both photometry and polarimetry are
on a par with those of the results obtained with the 6-m
telescope.

\section{CONCLUSIONS}
\label{S:con} The ``StoP'' instrument was put into operation in
the early 2020. We demonstrate that the instrument can achieve the
accuracy needed for the photometric and polarimetric study of
extragalactic objects. During the first half-year of its operation
significant scientific results were obtained including both those
reported in this paper and those submitted to other scientific
journals~\citep{M20}. Hence the instrument improves the efficiency
of observations and expands the range of observational tasks that
can be addressed with the 1-m Zeiss-1000 telescope of the Special
Astrophysical Observatory of the Russian Academy of Sciences.

\begin{acknowledgments}
We are grateful to V.~V.~Komarov and to the engineers of the
Observations Support Laboratory (OSL) of the Special Astrophysical
Observatory for technical assistance in carrying out observations
with the Zeiss-1000 telescope, to V.~R.~Amirkhanyan for the setup
of the electronic part of the instrument, and the staff of the OSL
for providing access to MMPP photometer-polarimeter.
\end{acknowledgments}

\section*{FUNDING}
This work was supported by the Russian Science Foundation (grant
no.~ 20-12-00030 ``Investigation of geometry and kinematics of
ionized gas in active galactic nuclei by polarimetry methods''.
Observations were conducted with the telescopes of the Special
Astrophysical Observatory of the Russian Academy of Sciences  with
the financial support of the Ministry of Science and Higher
Education of the Russian Federation (including agreement No.
05.619.21.0016, project ID RFMEFI61919X0016).

\section*{CONFLICT OF INTEREST}
The authors declare no conflict of interest.

\onecolumngrid
\begin{flushright}
\textit{ Translated by A.~Dambis}
\end{flushright}

\end{document}